\documentclass[epj]{svjour}
\usepackage{graphicx}
\usepackage{dcolumn}
\usepackage{bm}

\usepackage{braket}
\usepackage{multirow}
\usepackage{dsfont}
\usepackage{mathtools}
\usepackage{xcolor}
\usepackage[normalem]{ulem}

\begin{document}

\title{Chiralspin symmetry and  QCD at high temperature}
\author{L. Ya. Glozman  
}                     
\institute{Institute of Physics,  University of Graz, A--8010 Graz, Austria}
\date{Received: date / Revised version: date}
%

\abstract{
It has been found very recently on the lattice that
at high temperature at vanishing chemical potential QCD is increasingly $SU(2)_{CS}$
and $SU(2N_F)$ symmetric. We demonstrate that the  chemical potential
term in the QCD Lagrangian has precisely the same symmetry. Consequently
the QCD matter beyond the chiral restoration line at high temperature
on the $T-\mu$
plane is at least approximately $SU(2)_{CS}$ and $SU(2N_F)$ symmetric.
} 

\maketitle

\section{\label{sec:intro}Introduction}

A structure of the QCD phase diagram as well as a nature of the
strongly interacting matter in different regimes attracts enormous 
experimental and theoretical efforts. It is established in  QCD calculations 
on the lattice that there is a transition to the chirally symmetric regime at large temperatures and low densities, where the
quark condensate, an order parameter of $SU(N_F)_L \times SU(N_F)_R$ chiral
symmetry, vanishes. In addition there is a strong evidence that
above the critical temperature also the $U(1)_A$ symmetry gets restored 
\cite{Cossu:2013uua,Tomiya:2016jwr,Bazavov:2012qja}. Very recently $N_F=2$
lattice simulations with the domain-wall Dirac operator have demonstrated 
 emergence  of  $SU(2)_{CS}$
and $SU(4)$ symmetries \cite{Glozman:2014mka,Glozman:2015qva} at increasing temperature \cite{R}.
These symmetries have been observed earlier in dynamical lattice
simulations upon artificial truncation  of the near-zero
modes of the Dirac operator at zero temperature
\cite{Denissenya:2014poa,Denissenya:2014ywa,Denissenya:2015mqa,Denissenya:2015woa}. 
The $SU(2)_{CS} \supset U(1)_A$
and $SU(2N_F) \supset SU(N_F)_L \times SU(N_F)_R \times U(1)_A$  symmetries
are symmetries of the chromo-electric interaction in QCD. 
In addition to the chiral transformations the $SU(2)_{CS}$ and $SU(2N_F)$
rotations mix the left- and right-handed components of quark fields.
The chromo-magnetic
interaction as well as the quark kinetic term break these symmetries down
to $SU(N_F)_L \times SU(N_F)_R \times U(1)_A$.

Here we demonstrate that the quark chemical potential term in the
QCD Lagrangian is $SU(2)_{CS}$ and $SU(2N_F)$ symmetric, i.e. it
has the same symmetry as the confining chromo-electric interaction.
Consequently the quark chemical potential term can only impose
the $SU(2)_{CS}$ and $SU(2N_F)$
symmetries of confinement. This means that  QCD at high temperature
beyond the
chiral symmetry restoration line  on the $\mu -T$ plane, where the quark condensate vanishes, should have approximate $SU(2)_{CS}$ and $SU(2N_F)$
symmetries with increasing accuracy with temperature and chemical potential. 

\section{\label{sec:su4}$SU(2)_{CS}$ and $SU(2N_F)$ symmetries \cite{Glozman:2014mka,Glozman:2015qva}}

The  $SU(2)_{CS}$  chiralspin transformations,
defined in the Dirac spinor space  are 

\begin{equation}
\label{V-def}
  \Psi \rightarrow  \Psi^\prime = e^{i  {\bf{\varepsilon} \cdot \bf{\Sigma}}/{2}} \Psi  \; ,
\end{equation}
\noindent
with the following generators

\begin{equation}
\vec \Sigma = \{\gamma_k,-i \gamma_5\gamma_k,\gamma_5\},
\label{eq:su2cs_}
\end{equation}
\noindent
$k=1,2,3,4$. Different $k$ define different irreducible
representations of dim=2.
$U(1)_A$ is a subgroup of $SU(2)_{CS}$.
The $su(2)$ algebra
$[\Sigma_\alpha,\Sigma_\beta]=2i\epsilon^{\alpha\beta\gamma}\Sigma_\gamma$
is satisfied with any Euclidean gamma-matrix, obeying the following
anticommutation relations

\begin{equation}
\gamma_i\gamma_j + \gamma_j \gamma_i =
2\delta^{ij}; \qquad \gamma_5 = \gamma_1\gamma_2\gamma_3\gamma_4.
\label{eq:diracalgebra}
\end{equation}
\noindent
  The  $SU(2)_{CS}$ transformations mix the
left- and right-handed fermions. The free massless quark Lagrangian
does not have this symmetry.

An extension of the $SU(2)_{CS} \times SU(N_F)$ product
leads to a $SU(2N_F)$ group. This group has the chiral
symmetry of QCD $SU(N_F)_L \times SU(N_F)_R \times U(1)_A$ as a subgroup.
Its transformations and generators are given by
\begin{equation}
\label{W-def}
\Psi \rightarrow  \Psi^\prime = e^{i \bf{\epsilon} \cdot \bf{T}/2} \Psi\; ,
\end{equation}

\begin{align}
\{
(\tau_a \otimes \mathds{1}_D),
(\mathds{1}_F \otimes \Sigma_i),
(\tau_a \otimes \Sigma_i)
\}
\end{align}
where $\tau$  are flavour generators with flavour index $a$ and $i=1,2,3$ is the $SU(2)_{CS}$ index.

The fundamental
vector of $SU(2N_F)$ at $N_F=2$ is

\begin{equation}
\Psi =\begin{pmatrix} u_{\textsc{L}} \\ u_{\textsc{R}}  \\ d_{\textsc{L}}  \\ d_{\textsc{R}} \end{pmatrix}. 
\end{equation}
\noindent
The $SU(2N_F)$ transformations mix both flavour and chirality.

\section{Symmetries of different parts of the QCD Lagrangian
and $SU(2)_{CS}$, $SU(2N_F)$  emergence at high temperatures.}

The interaction of  quarks with the gluon field in Minkowski space-time
can be splitted into a temporal and a spatial part:

\begin{equation}
\label{Lagrangian}
 \overline{\Psi}   \gamma^{\mu} D_{\mu} \Psi = \overline{\Psi}   \gamma^0 D_0  \Psi 
  + \overline{\Psi}   \gamma^i D_i  \Psi\; . 
\end{equation}
The first (temporal) term includes an interaction of the color-octet
quark charge density 
$\bar \Psi (x)  \gamma^0 \vec \lambda \Psi(x) = \Psi (x)^\dagger \vec \lambda \Psi(x)$
with the chromo-electric  
part of the gluonic field ($\vec \lambda$ are color Gell-Mann matrices). It is invariant with respect to 
any unitary transformation that can be defined in the Dirac spinor space,
in particular it is invariant under
the chiral transformations, 
the $SU(2)_{CS}$ transformations (1) as well as the  transformations (4). 
 
 The spatial part contains a quark kinetic term
and  an
interaction of the chromo-magnetic field with the color-octet
spatial current density.
This spatial part is invariant only under chiral
$SU(N_F)_L \times SU(N_F)_R \times U(1)_A$ transformations
and does not admit higher $SU(2)_{CS}$ and $SU(2N_F)$ symmetries.  
Consequently the QCD Lagrangian has, in the chiral limit, only the 
 $U(N_F)_L \times U(N_F)_R$
chiral symmetry.

It was found on the lattice with  chirally-invariant fermions
in $N_F=2$ dynamical simulations that truncation of the near-zero modes of the Dirac operator results in emergence of the  $SU(2)_{CS}$ and $SU(4)$
symmetries in hadrons \cite{Denissenya:2014poa,Denissenya:2014ywa,Denissenya:2015mqa,Denissenya:2015woa}.

The  emergence of the $SU(2)_{CS}$ and $SU(4)$
symmetries upon truncation of the lowest modes of the Dirac operator
means that the effect
of the chromo-magnetic interaction in QCD is located exclusively in
the near-zero modes. At the same time the confining chromo-electric interaction,
which is $SU(2)_{CS}$- and $SU(4)$-symmetric, is
distributed among all modes of the Dirac operator.

To conclude, the low-lying modes of the Dirac operator are
responsible not only for chiral symmetry breaking, as it is seen from the 
Banks-Casher relation \cite{BC},
but also
for the $SU(2)_{CS}$ and $SU(4)$ breaking via the magnetic effects.
The magnetic effects are linked exclusively to the near-zero
modes. 

Given this insight one could expect emergence of the $SU(2)_{CS}$ and $SU(4)$
symmetries at high temperatures, because at high temperature the
near-zero modes of the Dirac opperator are suppressed. This expectation
 has been confirmed very recently in lattice simulations with chiral fermions \cite{R}.
It was found that indeed above the critical temperature at vanishing chemical potential the approximate
$SU(2)_{CS}$ and $SU(4)$ symmetries are seen in spatial correlation
functions and by increasing the temperature the $SU(2)_{CS}$ and $SU(4)$ breaking effects decrease rapidly; at the highest available temperature
380 MeV these breaking effects are at the level of 5\%.

\section{Symmetries of the quark  chemical potential.}

Will a non-zero  chemical
potential break this symmetry?
Consider the quark part of the Euclidean QCD action at a temperature $T=1/\beta$ in a medium
with the quark chemical potential $\mu$:

\begin{equation}
S = \int_{0}^{\beta} d\tau \int d^3x
\overline{\Psi}  [ \gamma_{\mu} D_{\mu} + \mu \gamma_4 ] \Psi,
\end{equation}

\noindent
where $\Psi$ and $\overline{\Psi}$ are independent integration
variables. The field $\overline{\Psi}$ is defined such that
it transforms like $ \Psi^\dagger \gamma_4$, i.e. like Minkowskian $\overline{\Psi}$.

This means that the quark chemical potential term

\begin{equation}
\mu \overline{\Psi}  \gamma_4 \Psi 
\end{equation}
\noindent
transforms under chiral,
 $SU(2)_{CS}$ and $SU(2N_F)$ transformations as 

\begin{equation}
\mu {\Psi}^\dagger \Psi, 
\end{equation}

\noindent
i.e. it is invariant under all these
unitary groups. In other words, the dense QCD matter not only
does not break the $SU(2)_{CS}$ and $SU(2N_F)$ symmetries, but in a sense
imposes them since the chemical potential $\mu$ is an external
parameter that can be arbitrary large. The chemical potential term is a
color-singlet. Consequently this term can only reinforce the
$SU(2)_{CS}$ and $SU(2N_F)$ symmetries at high temperature and zero
chemical potential arising from the chromo-electric color-octet term
and a compensation is impossible.

\begin{figure}
\centering 
\includegraphics[width=0.9\linewidth]{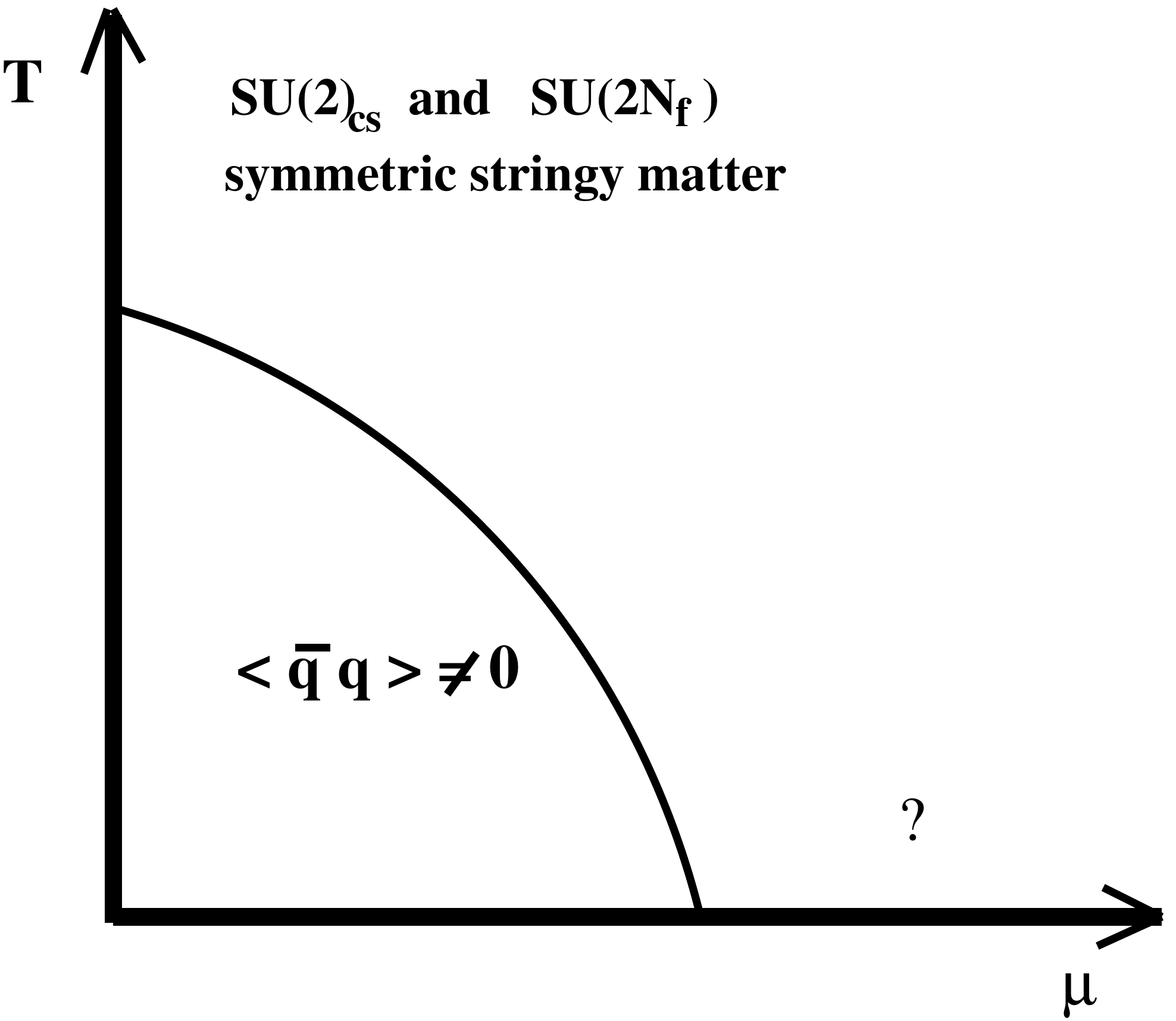}
\caption{A sketch of the QCD phase diagram at high temperatures 
and chemical potentials.} 
\end{figure}

We conclude that at high temperature $T \sim 400$ MeV at any chemical potential the QCD matter is
approximately $SU(2)_{CS}$- and $SU(4)$-symmetric. 
The $SU(2)_{CS}$ and $SU(4)$ symmetries
emerge due to yet unknown microscopic dynamics. This dynamics
suppresses (screens) the chromo-magnetic field while the chromo-electric
interaction between  quarks is still active.\footnote{A plausible 
microscopic explanation
of this phenomenon could be related to suppression at high T of the 
local topological fluctuations of the gluonic field, like instantons,
monopols etc. According to the Atiyah-Singer theorem difference of the
number of the left- and right-handed zero modes of the Dirac operator
is related to the topological charge $Q$ of the gauge configuration.
Consequently with $|Q| \geq 1$ amount of the right- and left-handed
zero modes is not equal which manifestly  breaks the $SU(2)_{CS}$
symmetry since the $SU(2)_{CS}$ transformations mix the left- and
right-handed components of quarks. The topological configurations
contain the chromo-magnetic field. What would be exact zero modes
become the near-zero modes of the  Dirac operator in the global
gauge configuration that contain local topological fluctuations,
like in the Shuryak-Diakonov-Petrov theory of chiral symmetry breaking
in the instanton liquid. Consequently all effects of the chromo-magnetic
field are localised in the near-zero modes, while confining chromo-electric
field is distributed among all modes. At $T > T_C$ the local toplogical
fluctuations are melt what leads first to restoration of chiral symmetry
and then to $SU(2)_{CS}$ emergence.}

 The elementary
objects in the high temperature QCD matter are chiral quarks connected
by the chromo-electric field, without any magnetic effects, a kind
of a string \cite{GS}. These objects cannot be described
as bound states in some nonrelativistic potential. With the nonrelativistic Schr\"{o}dinger equation appearance of chiral as well as of $SU(2)_{CS}$ and $SU(4)$ symmetries is impossible. Consequently the QCD matter at
high temperature and low chemical potential could be named a "stringy
matter", see Fig. 1.

\section{Conclusions}

The main new insight of this short note is that the approximate
$SU(2)_{CS}$ and $SU(2N_F)$ symmetries emerge at a temperature $\sim 2 T_c$
 on the $T -\mu$ phase diagram and their breaking decreases 
with increased chemical potential. So we can consider the QCD matter at
these temperatures  as at least approximately 
$SU(2)_{CS}$- and $SU(2N_F)$-symmetric. These symmetries rule out
the asymptotically free deconfined quarks:  free quarks are
incompatible with these symmetries. Note that these symmetries cannot be
obtained in perturbation theory which relies on a symmetry of a free
Dirac equation, i.e. on chiral symmetry. The elementary objects in
the QCD matter at these temperatures are chiral quarks connected by the
chromo-electric field. Such a matter is not a 
quark-gluon plasma (the plasma notion is defined in physics as a system
of free charges with Debye screening of the electric field) and could be
more adequately named as a stringy fluid.

\bigskip
I am thankful to M. Chernodub who provoked this analysis
by asking the author about the fate of the chiralspin symmetry in a
dense medium.
Partial support from the Austrian Science Fund (FWF) through the grant
 P26627-N27 is acknowledged.



\begin{thebibliography}{30}

\bibitem{Cossu:2013uua} 
  G.~Cossu, S.~Aoki, H.~Fukaya, S.~Hashimoto, T.~Kaneko, H.~Matsufuru and J.~I.~Noaki,
  Phys.\ Rev.\ D {\bf 87}, (2013) 114514. 
  Erratum: [Phys.\ Rev.\ D {\bf 88}, (2013) 019901].

\bibitem{Tomiya:2016jwr}
  A.~Tomiya, G.~Cossu, S.~Aoki, H.~Fukaya, S.~Hashimoto, T.~Kaneko and J.~Noaki, Phys.\ Rev.\ D {\bf 96}, (2017) 034509.

\bibitem{Bazavov:2012qja}
  A.~Bazavov {\it et al.} [HotQCD Collaboration],
  Phys.\ Rev.\ D {\bf 86}, (2012) 094503.
 
 \bibitem{Glozman:2014mka} 
  L.~Y.~Glozman,
  Eur.\ Phys.\ J.\ A {\bf 51}, (2015) 27.

\bibitem{Glozman:2015qva}
  L.~Y.~Glozman and M.~Pak,
  Phys.\ Rev.\ D {\bf 92},(2015) 016001.

\bibitem{R} 
  C.~Rohrhofer, Y.~Aoki, G.~Cossu, H.~Fukaya, L.~Y.~Glozman, S.~Hashimoto, C.~B.~Lang and S.~Prelovsek,
  Phys.\ Rev.\ D {\bf 96},(2017) 094501.

\bibitem{Denissenya:2014poa} 
  M.~Denissenya, L.~Y.~Glozman and C.~B.~Lang,
  Phys.\ Rev.\ D {\bf 89}, (2014) 077502.

\bibitem{Denissenya:2014ywa} 
  M.~Denissenya, L.~Y.~Glozman and C.~B.~Lang,
  Phys.\ Rev.\ D {\bf 91}, (2015) 034505.

\bibitem{Denissenya:2015mqa} 
  M.~Denissenya, L.~Y.~Glozman and M.~Pak,
  Phys.\ Rev.\ D {\bf 91}, (2015) 114512.

\bibitem{Denissenya:2015woa} 
  M.~Denissenya, L.~Y.~Glozman and M.~Pak,
  Phys.\ Rev.\ D {\bf 92}, (2015) 074508
  Erratum: [Phys.\ Rev.\ D {\bf 92}, (2015) 099902].


\bibitem{BC} 
T.~Banks and A.~Casher,
  Nucl.\ Phys.\ B {\bf 169} (1980) 103.

\bibitem{GS} 
 L.~Y.~Glozman,
  Phys.\ Lett.\ B {\bf 541}, (2002) 115.




\end{thebibliography}
\end{document}